\documentclass[10pt,a4paper,oneside]{article}
\usepackage[a4paper, margin=1.3in, total={210mm, 297mm}]{geometry}
\usepackage{apacite}    
\usepackage[utf8]{inputenc} 
\usepackage{fancyhdr}
\usepackage{chngcntr}
\usepackage{mathtools,amssymb}
\usepackage{dsfont}
\usepackage{caption}
\usepackage{subcaption}
\usepackage{float}
\usepackage{tikz}
\usetikzlibrary{datavisualization.formats.functions}
\usepackage{pgfplots}
\usepackage[inline]{enumitem}
\usepackage{amsthm}

\begin{document} 
	
	\begin{center}
		{Confidence interval {for the sensitive fraction}\\ in Item Count Technique model\\}
		%date{\today}
		
		\bigskip
		
		Stanis{\l}aw Jaworski\\
		Wojciech Zieli{\'n}ski\\
		Warsaw University of Life Sciences (Poland)\\
		e-mail: stanislaw$\_$jaworski@sggw.edu.pl\\
		e-mail: wojciech$\_$zielinski@sggw.edu.pl\\	
		
	\end{center}
	
    \vfill\eject
	
\centerline{\bf Abstract}

The problem is in the estimation of the fraction of population with a sensitive characteristic. We consider the Item Count Technique { an indirect method of questioning designed
	to protect respondents' privacy. The exact confidence interval for the sensitive fraction is
	constructed. The length of the proposed CI depends on both the given parameter of the
	model and the sample size. For these CI the model's parameter is established in relation
	to the provided level of the privacy protection of the interviewee. The optimal sample size
	for obtaining a CI of a given length is discussed in the context.

	Keywords: sensitive questions, Item Count Technique, exact confidence interval

	2010 Mathematics Subject Classification: 62F25, 62P20

\bigskip

\noindent{\bf 1. Introduction}
	
	The problem is in the estimation of the percentage of population with some sensitive characteristic, e.g. related to corruption, tax frauds, sexual abuse, illegal work (black market), drug
	uses, violence against children or other socially stigmatizing behaviours.
	Mathematicaly, let $Z$ be a random variable such that
	$$P\{Z = 1\}= \pi = 1 -P\{Z = 0\}.$$
	This variable takes on the value $1$ when the answer to the sensitive question is YES and
	the value $0$ otherwise. The number $\pi\in(0; 1)$ is the probability of the positive answer to
	the sensitive question, i.e. $\pi\cdot100\%$ is the percentage of interest. We want to estimate the
	probability $\pi$ and to construct a confidence interval for $\pi$.
	
	The difficulty is that variable $Z$ is usually not directly observed. In survey questionnaires
	we rarely ask the sensitive question directly. Instead we use several techniques designed to
	protect respondents' privacy. Answers to the sensitive question are ``masked'' through asking
	a ``neutral'' question. It is assumed that the ``neutral'' question is independent from the
	sensitive question. In a questionnaire two questions are asked: sensitive and neutral. But only
	one answer is registered and the interviewer does not know which of the two questions the
	interviewee answered.
	
	The first method of obscuring the answer to a sensitive question was proposed by (Warner,
	1965). This method consists in randomization of the answers. The randomization is done by
	the respondent and the interviewer does not know which of the two questions the interviewee
	answered. This model was extended in different ways (Horvitz, Shah, \& Simmons, 1967;
	Greenberg, Abul-Ela, \& Horvitz, 1969; Raghavarao, 1978; Franklin, 1989; Arnab, Shangodoyin,
	\& Arcos, 2019; Arnab, 1990, 1996; Kuk, 1990; Rueda, Cobo, \& Arcos, 2015).
	
	Tian, Yu, Tang, and Geng (2007) proposed a nonrandomized response model (NRR). Their
	idea consists in asking two questions simultaneously: one sensitive and one neutral. This model
	was extended to other, similar approaches: (Yu, Tian, \& Tang, 2008; Tan, Tian, \& Tang, 2009;
	Tian, 2014).	
	
	In what follows we consider the Item Count Technique (ICT) in a version proposed in
	(Kowalczyk, Niemiro, \& Wieczorkowski, 2021) and described in section 3. We assume that
	the answer to the neutral question is a count variable, i.e. a random variable with nonnegative
	integer values (whilst the sensitive answer is a binary variable).
	
	Unfortunately, the problem of constructing a confidence intervals (CI's) for $\pi$ was considered rather rarely. Moreover, the proposed CI's are asymptotic. Their actual coverage
	probability may fall below the nominal confidence level. Consequently, they do not fulfil the
	basic requirement for confidence intervals (c.f. Neyman, 1934, p. 562). In what follows we
	propose a construction of finite sample size CI in the ICT model.
	
	In section 2 we describe the method for construction of an exact CI for $\pi$. In section 3 we
	discuss the problem of optimal sample size with respect to both privacy protection an accuracy
	of estimation. In section 4 we revisit the construction of asymptotic confidence interval and
	discuss the coverage probability in this context.   In section 5 some concluding remarks are given.

\bigskip	
\vfill\eject

\noindent{\bf 2. Confidence interval in ICT model}

%\label{sec: exact}\input{Sec/exact-intro}
In our considerations, we assume that the sample is randomly drawn from the
population and randomly divided into two treatment groups. Let us denote by
$Z$ a sensitive variable under study, $Z\sim Bernoulli(\pi)$; and by $X$ a 
count control variable. Thus we have
$$Y=
\begin{cases}
	X - Z,&\hbox{in the $1^{st}$ sample};\\
	X + Z,&\hbox{in the $2^{nd}$ sample}\\
\end{cases}.
$$
We assume that $X$ and $Z$ are independent, $Y$ is observed, $X$ and $Z$ are hidden (missing) variables. Throughout the paper, we assume ``no design effect'' (Assumption 1 in Blair \& Imai, 2012) and ``no liars assumption'' (Assumption 2 in Blair \& Imai, 2012), that is, we treat $Z$ as the truthful
answer. The whole sample consists of {$n=n_1+n_2$} independent variables
$Y_{11},\ldots, Y_{1,n_1}$ and $Y_{21},\ldots,Y_{2,n_2}$, where $Y_{1i}=X_{1i}-Z_{1i}$ for $i=1,\ldots,n_1$ and $Y_{2i}=X_{2i}+Z_{2i}$ for $i=1,\ldots,n_2$.

Consider two random variables: 
$$\sum_{i=1}^{n_2}Y_{2i}=\sum_{i=1}^{n_2}X_{2i}+\sum_{i=1}^{n_2}Z_{2i},$$
$$\sum_{i=1}^{n_1}Y_{1i}=\sum_{i=1}^{n_1}X_{1i}-\sum_{i=1}^{n_1}Z_{1i}.$$
Note that $\sum_{i=1}^{n_2}Z_{2i}\sim Bin(n_2,\pi)$, $\sum_{i=1}^{n_1}Z_{1i}\sim Bin(n_1,\pi)$ and these two sums are 
independent ($Bin(n,\pi)$ denotes binomial distribution with parameters $n$ and $\pi$).
Since the random variable $$\frac{1}{n_1}\sum_{i=1}^{n_2}Y_{2i}-\frac{1}{n_2}\sum_{i=1}^{n_1}Y_{1i}$$ can be considered an unbiased estimator of $\pi$, our construction of the exact confidence interval is based on the statistic 
$$ \sum_{i=1}^{n_2}Y_{2i}-\sum_{i=1}^{n_1}Y_{1i}= \left(\sum_{i=1}^{n_2}X_{2i}-\sum_{i=1}^{n_1}X_{1i}\right)+\left(\sum_{i=1}^{n_2}Z_{2i}+\sum_{i=1}^{n_1}Z_{1i}\right).$$
This statistic is a sum of two independent random variables, say $\xi+\eta$, where $\eta\sim Bin(n_1+n_2,\pi)$. Hence the cumulative distribution function (CDF) of this statistic is of the form
\begin{equation}\label{D}
	F_\pi(z)=P_\pi\left\{\xi+\eta\leq z\right\}=\sum_{k=0}^{n}P\left\{\xi\leq z-k\right\}P_\pi\left\{\eta=k\right\}
\end{equation}

It is evident that this family of distributions is stochastically ordered with respect to $\pi$. For a given $z$, a CI for $\pi$ at the confidence level $\gamma$ can be obtained by solving the following problem:
\begin{equation}
	\label{P}
	\begin{cases}
		\pi_L(z;\gamma)=\arg\inf_\pi F_\pi(z)\geq \frac{1+\gamma}{2}\\
		\pi_R(z;\gamma)=\arg\sup_\pi F_\pi(z^-)< \frac{1-\gamma}{2}\\
	\end{cases}
\end{equation}

where $F_\pi(z^-)=P_\pi\left\{\xi+\eta< z\right\}$.

In what follows we consider the case of a Poisson-distributed response to the neutral question. Thus, we assume that the random variable $X$ is distributed as Poisson $Po(\lambda)$ with mean value $\lambda>0$. Then 
$$\sum_{i=1}^{n_1}X_{1i}\sim Po(n_1\lambda); \qquad\sum_{i=1}^{n_2}X_{2i}\sim Po(n_2\lambda).$$

It is known that if random variable $\xi_1$ is distributed as Poisson with mean value $n_1\lambda$ and random variable $\xi_2$ is distributed as Poisson with parameter $n_2\lambda$ and those random variables are independent then $\xi=\xi_2-\xi_1$ is distributed as Skellam with the probability mass function
$$g(k)=P(\xi=k)=e^{-(n_2\lambda+n_1\lambda)}\left(\frac{n_2\lambda}{n_1\lambda}\right)^{k/2}I_{|k|}(2\sqrt{n_2\lambda n_1\lambda}), \hbox{ for integer $k$},$$
where $I_{|k|}(z)$ is modified Bessel function of the first kind. The CDF equals
$$G(x)=P(\xi\leq x)=1-Q_{-\lfloor x\rfloor }\left(\sqrt{n_2\lambda},\sqrt{n_1\lambda}\right),$$
where $Q_{x}(a,b)$ is a Macrum function. Procedures for numerically computing
the above functions $g$ and $G$ are easily available in mathematical or statistical packages (for example R project). Therefore the CDF $F_\pi$ in equation $(\ref{D})$ can be evaluated as 
$$\begin{aligned}
	F_{\pi}(z)&=
	&\sum_{k=0}^{n}\left(1-Q_{-\lfloor z-k\rfloor }\left(\sqrt{n_2\lambda},\sqrt{n_1\lambda}\right)\right)\binom{n}{k}\pi^k(1-\pi)^{n-k}.\\
\end{aligned}
$$
The lower and upper bounds of our confidence interval given by $2$ can be determined by solving two nonlinear equations with a single unknown. Unfortunately, the bounds of the CI can only be found numerically.

\bigskip

In what follows we take $n_1=n_2=n/2$. 
\begin{figure}[H]
	\begin{center}
		\includegraphics[scale=0.6]{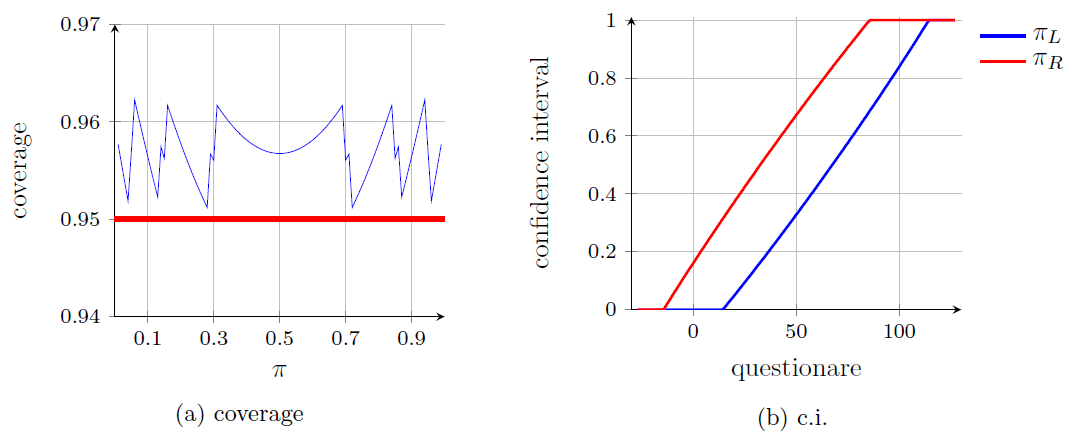}
	\end{center}
\end{figure}
\centerline{Figure 1. $n_1=n_2=50, n=100, \lambda=0.5$}

\bigskip
\noindent{\bf 3. Optimisation}
%\label{sec: optimisation}\input{Sec/optim-intro}

The issue of selecting the optimal sample size in studies involving sensitive questions has been addressed by many authors.
In their studies, two approaches to this problem can be distinguished.
The first approach (e.g. Tian, Tang, Liu, Tan, \& Tang, 2011) involves conducting precision and power analyses for one-sided and two-sided tests, focusing on the hypothesis $H_0:~\pi=\pi_0$, where $\pi$ denotes the population proportion with the sensitive characteristic and $\pi_0$ is a predefined reference value. Sample size determination is driven by controlling type I and type II error rates of the tests.

The second approach (e.g. Qiu, Zou, \& Tang, 2014; Qiu, Tang, Tao, \& Wong, 2022) emphasizes probabilistic control over the maximum length of the confidence interval for $\pi$. However, the formulas used to determine the optimal sample size depend on the true, unknown value of $\pi$, which complicates their direct application. Additionally, both approaches have two main drawbacks: they rely on approximate statistics and, more critically, do not ensure control over the level of privacy protection for respondents. This protection is influenced not only by the unknown parameter $\pi$ but also by the distribution function of respondents' answers to neutral question. The greater the variability of this distribution, the higher the privacy protection for the respondent, but also the lower the precision of estimating the proportion of individuals with the sensitive trait. In our study, we assume a Poisson distribution of respondent's answers to neutral question. The variance of this distribution is equal to $\lambda$. Therefore, we expect that as $\lambda$ increases, privacy protection improves, while estimation precision decreases. Since privacy protection is a priority, it is sensible to first determine the smallest value of $\lambda$ that maintains this protection at an appropriately high level, and then to establish the sample size needed for accurate estimation. We can encounter such an approach in work by Jaworski and Zieli{\'n}ski (2023).

\bigskip
\noindent{\bf 3.1 Choice of $\lambda$}

Tan et al. (2009) introduced a notion of the degree of privacy protection  through probabilities $$P_{\pi,\lambda}\left\{Z=1|Y=y\right\},$$ 
where $Y$ is an answer of the respondent. The privacy protection is a probability of quessing positive answer ($Z=1$) for sensivity question knowing the answer $Y$ to the whole questionare ($Y_1$ for the first group, $Y_2$ for the second group). This probability should be appropriately small.
In the problem of sensitive questions it is important to make the interviewee safety that her/his positive answer to the sensitive question will not be revealed. Assume that privacy protection is smaller then given probability small number $\delta$.

In ICT model for the first group (with subtraction) we have
\begin{equation}
	\begin{aligned}
		P_{\pi,\lambda}\left\{Z_1=1|Y_1=y_1\right\}&=\frac{P_{\pi,\lambda}\left\{Z_1=1\&Y_1=y_1\right\}}{P_{\pi,\lambda}\left\{Y_1=y_1\right\}}=\frac{1}{1+\frac{y_1+1}{\lambda}(\frac{1}{\pi}-1)}\\
	\end{aligned}
\end{equation}
for $y_1=-1,0,1,\ldots$.

For the second group (with addition)
\begin{equation}\begin{aligned}
		P_{\pi,\lambda}\left\{Z_2=1|Y_2=y_2\right\}&=\frac{P_{\pi,\lambda}\left\{Z_2=1\&Y_2=y_2\right\}}{P_{\pi,\lambda}\left\{Y_2=y_2\right\}}=\frac{1}{1+\frac{\lambda}{y_2}(\frac{1}{\pi}-1)}\\
	\end{aligned}
\end{equation}
for $y_2=1,2,3,\ldots$ and $P_{\pi,\lambda}\left\{Z=1|Y_2=0\right\}=0.$

Let $\delta$ and $\tau$ be given values from the interval $(0,1)$. W intend to find such a $\lambda$ that for every $\pi\leq \pi_0$ holds
\begin{equation}\label{ine: general}
	\begin{cases} P_{\pi,\lambda}\big\{ \{P_{\pi,\lambda}\left\{Z_1=1|Y_1\right\}\leq\tau\} \big\}\geq\delta\\
		P_{\pi,\lambda}\big\{ \{P_{\pi,\lambda}\left\{Z_2=1|Y_2\right\}\leq\tau\} \big\}\geq\delta
	\end{cases},
\end{equation}
where $\pi_0$ is a known value.\par 
The value $\pi_0$ reflects our prior knowledge  about the parameter $\pi$. We consider a rare sesitive phenomena, i.e. we assume that $\pi\leq\pi_0$ for given $\pi_0\leq 0.5$, where probability $\pi_0$ is given in advance.  For example, approximately 0.36\% of the U.S. population lives with HIV, and the number of new HIV infections and AIDS cases has been steadily declining due to advances in treatment and prevention. Given this information, and considering the population, we may assume $\pi_0=0.0036$ when re-estimating the percentage of people with AIDS.\par
In this approach, $\tau=0.5$ means that even if it were possible to link the survey result to a specific respondent, the reliability of the assessment that they belong to a sensitive group would not exceed 50\%.\par
Let  us consider the probabilities in  (\ref{ine: general}).

\begin{equation}
	\begin{aligned}
		P_{\pi,\lambda}\big\{ \{P_{\pi,\lambda}\left\{Z_1=1|Y_1\right\}\leq\tau\} \big\}=P_{\pi,\lambda}\left\{\frac{1}{1+\frac{Y_1+1}{\lambda}(\frac{1}{\pi}-1)}\leq\tau\right\}
		=P_{\pi,\lambda}\left\{Y_1\geq \lambda\frac{\frac{1}{\tau}-1}{\frac{1}{\pi}-1}-1\right\}
	\end{aligned}
\end{equation}

\begin{equation}
	\begin{aligned}
		P_{\pi,\lambda}\big\{ \{P_{\pi,\lambda}\left\{Z_2=1|Y_1\right\}\leq\tau\} \big\}=P_{\pi,\lambda}\left\{\frac{1}{1+\frac{\lambda}{Y_2}(\frac{1}{\pi}-1)}\leq\tau\right\}		=P_{\pi,\lambda}\left\{Y_2\leq \lambda\frac{\frac{1}{\pi}-1}{\frac{1}{\tau}-1}\right\}
	\end{aligned}
\end{equation}
Let $V=\lambda\frac{\frac{1}{\tau}-1}{\frac{1}{\pi}-1}-1$, $W=\lambda\frac{\frac{1}{\pi}-1}{\frac{1}{\tau}-1}$ and  $F_{\lambda}$, $P_{\lambda}$ denote the cdf and pdf of $Po(\lambda)$ accordingly.
\begin{equation}
	\begin{aligned}
		P_{\pi,\lambda}\left\{Y_2\leq \lambda\frac{\frac{1}{\pi}-1}{\frac{1}{\tau}-1}\right\}&=P_{\pi,\lambda}\left\{Y_2\leq W\right\}\\
		&= F_\lambda(\lfloor W\rfloor-1)+(1-\pi)P_\lambda(\lfloor W\rfloor)\\
	\end{aligned}
\end{equation}

\begin{equation}
	\begin{aligned}
		P_{\pi,\lambda}\left\{Y_1\geq \lambda\frac{\frac{1}{\tau}-1}{\frac{1}{\pi}-1}-1\right\}&=	P_{\pi,\lambda}\{Y_1\geq V\}\\
		&=1-\big( F_{\lambda}(\lceil V\rceil-1)+\pi P_{\lambda}(\lceil V\rceil)\big),
	\end{aligned}
\end{equation}
where $\lfloor\cdot\rfloor$ denotes the floor function and $\lceil\cdot \rceil$ the ceiling function. \par
We can reformulate the general criterion in the following way.
Find $\lambda$ such that for every $\pi\leq\pi_0$ holds
\begin{equation}\label{ine: generaltwo}
	\begin{cases} F_{\lambda}(\lfloor W\rfloor-1)+\pi P_{\lambda}(\lfloor W\rfloor)\geq\delta\\
		1-\big(F_{\lambda}(\lceil V\rceil-1)+\pi P_{\lambda}(\lceil V\rceil)\big)\geq \delta
	\end{cases}.
\end{equation}
The optimal choice of $\lambda$ depends on given $\delta,\tau$ and known $\pi_0$. The value $\lambda$ can not be too small, as the privacy criterion should hold. It is ilustrated in Fig. 2, where the smallest $\lambda$ for $\delta=0.95,\ \tau=0.05$ and $\pi_0=0.3$ is equal to $\lambda_{min}=5.364$. More examples are given in Tab. 1.

\begin{figure}[H]
	\begin{center}
		\includegraphics[scale=0.6]{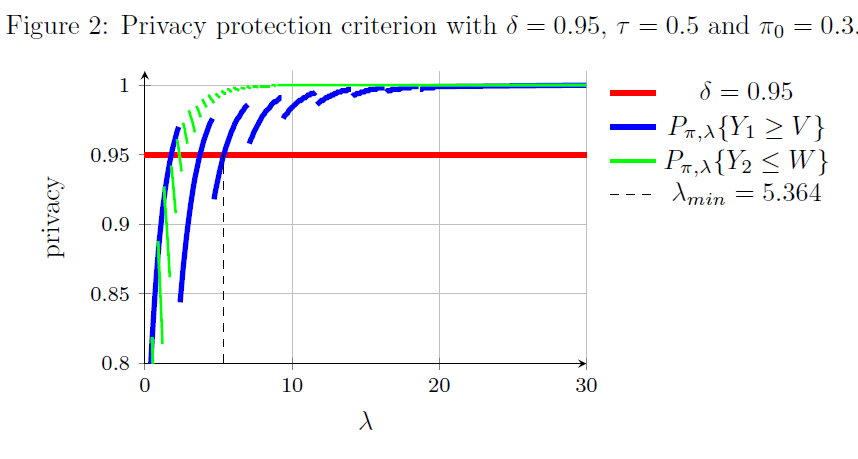}
	\end{center}
\end{figure}

\begin{figure}[H]
	\begin{center}
		\includegraphics[scale=0.6]{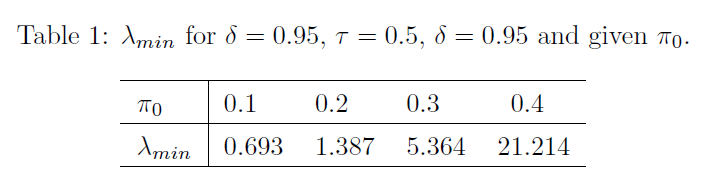}
	\end{center}
\end{figure}

\bigskip
\noindent{\bf 3.2 Sample size}
%\label{subsec: nchoice}\input{Sec/optim-n-choise}

In practical applications we are interested in short confidence intervals. The length $l(Y,\gamma;\lambda,n_1,n_2)$ of the CI is a random variable, where $Y= \sum_{i=1}^{n_2}Y_{2i}-\sum_{i=1}^{n_1}Y_{1i}$. It depends on $\lambda$, $n_1,\ n_2$. A sample of size $n$ is divided randomly into two parts of sizes $n_1$ and $n_2$ respectively. We assume subsamples are of equal sizes, i.e. assume $n_1=n_2=n/2$. Hence the length of the CI we will denote by $l(Y,\gamma;\lambda,n)$.

Minimizing the length of CI for observed $y$ relays on changing probabilities of over- and underestimation. This method was extensively discussed in Zieli{\'n}ski (2010, 2017, 2022). To obtain the shortest CI in Neyman sense, i.e. controlling the probability of coverage, randomization is needed. 

\begin{figure}[h]
	\begin{center}
		\includegraphics[scale=0.5]{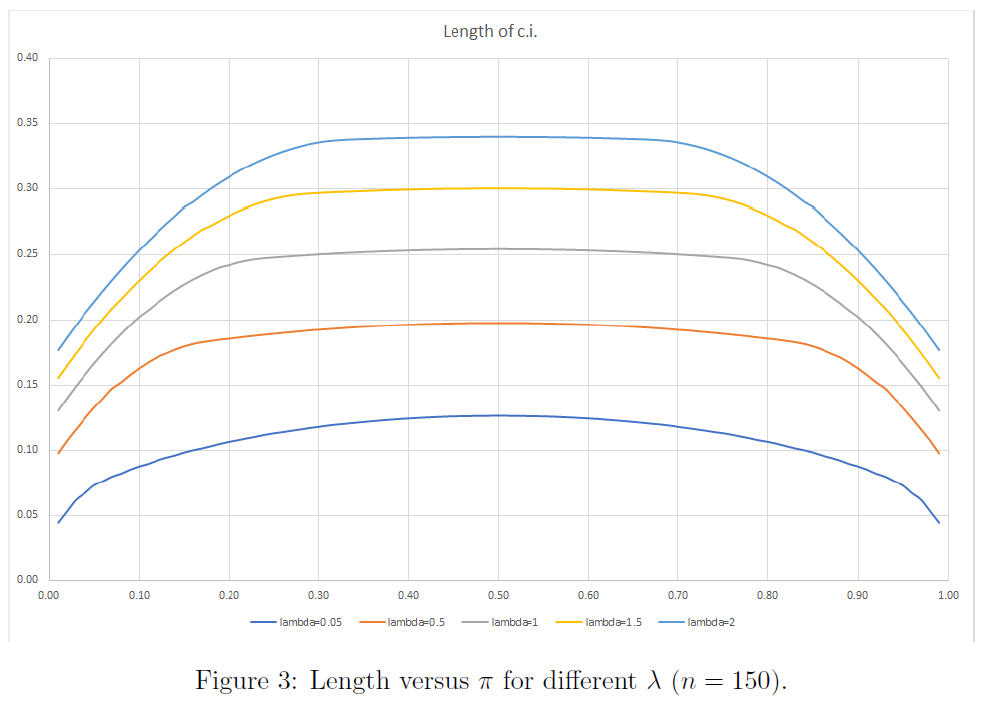}
	\end{center}
\end{figure}

The analysis of the length shows that (Figure 3)%, \ref{graph:lam1})
\begin{itemize}
	\item length increasis as $\lambda$ increasis
	\item length is maximal for $\pi=0.5$
	%\item length decreasis as $n$ grows,
\end{itemize}
As we expected, the best estimation precision can be achieved with the smallest possible value of $\lambda$. Once this value is established to meet the privacy criterion, we can proceed to determine the sample size based on the length of the confidence interval. 

Let us consider the length $l(Y,\gamma;\lambda,n)$ of the exact confidence interval $2$. We are interested in determing sample size giving the CI of a given length. We consider two approaches to the problem. In the first approach, our objective is to ensure that the average length of the CI covering the estimated value of $\pi$ is less than a given $d$. In the second approach, our goal is to ensure that the length of at least $\Lambda\%$ of the CI is less than the specified $d$.

\def\Jdn{\mathds{1}}
Considered  criterions may be formalized in the following way:
\begin{enumerate}
	\item for given $d$ find  minimal sample size $n$ such that $E{_\pi}l(Y,\gamma;\lambda,n)\leq d$ for every $\pi\leq\pi_0$
	\item for given $d$ and $\Lambda$  find  minimal sample size $n$ that $P_{\pi}\left\{l(Y,\gamma;\lambda,n)\leq d\right\}\geq \Lambda/\gamma$ for every $\pi\leq\pi_0$,
\end{enumerate}
where  $$E{_\pi}l(Y,\gamma;\lambda,n)=\sum_{y} l(y,\gamma;\lambda,n)P_\pi\{Y=y\}
\Jdn_{(\pi_L(y),\pi_R(y))}(\pi)$$ and $$P_{\pi}\left\{l(Y,\gamma;\lambda,n)\leq d\right\}=\sum_{y} P_\pi\{Y=y\}\Jdn(l(y,\gamma;\lambda,n)\leq d)\Jdn_{(\pi_L(y),\pi_R(y))}(\pi)$$
We assume that $\Lambda\in(0,1)$. Note that $\Lambda$ has been divided by $\gamma$. Since for sufficiently large $n$, the probability $P_{\pi}\left\{l(Y,\gamma;\lambda,n)\leq d\right\}$  approximately equals $\gamma$, it is convenient to normalize the right side of the inequality from the second criterion.

Unfortunately, those problems may be solved only numerically. %Some solutions will be presented in the next chapter.

\bigskip
\noindent{\bf 3.3. Examples}
%\label{sec: examples}\input{Sec/examples}

\bigskip
\noindent{\bf Example 3.1.}
We assume that confidence level $\delta=0.95$, $\pi\leq \pi_0=0.1$ and $\tau=0.5$ (according to the privacy protection criterion). Therefore, optimal $\lambda=\lambda_{min}=0.693$ (see Tab. 1).  Our goal is to determine the optimal sample size according to the first criterion. We want the expected length of the confidence interval not to exceed a given value $d$. For this purpose, we determine $\sup_{\pi\leq \pi_0}E_{\pi}l(Y,\gamma;\lambda_{min},n)$ (Tab. 2). According to the given calculations in Tab.~2 if $d=0.06$, optimal $n$ is approximately equal to $2\cdot 1550=3100$. If $d=0.05$, optimal $n$ is approximately equal to $2\cdot 2250=4500$.
\begin{figure}[h]
	\begin{center}
		\includegraphics[scale=0.5]{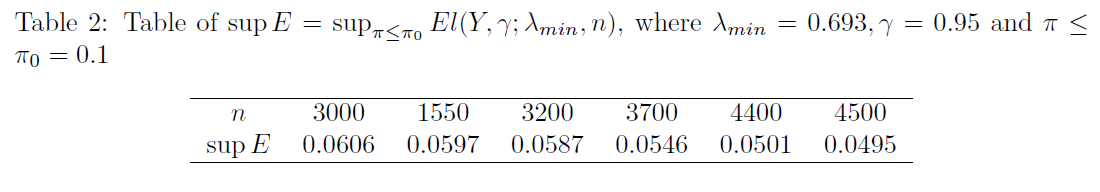}
	\end{center}
\end{figure}

%\bigskip
\noindent{\bf Example 3.2.}
Assume that  confidence level $\delta=0.95$,  $\pi_0=0.2$ and $\tau=0.5$. According to the privacy protection criterion the optimal $\lambda=\lambda_{min}=1.387$. Consider the length of CI that is  $l(y,\gamma;\lambda_{min},n)$ with respect to $y$. Typical shape of this mapping is depicted in Fig. 4. The graph of the confidence interval length is relatively flat. Hence, it seems that to approximate the optimal sample size, it is sufficient to observe the length of the interval. It is not true. For example, the maximum expected length of the confidence interval for $n=6000$ is $\sup_{\pi \leq\pi_0} El(Y, 0.95; \lambda_{min}, 6000) = 0.06$. 
Thus, $n=6000$ is the optimal sample size according to the first criterion of optimality with the given $d=0.06$. Note that for $n=6000$, the maximum length of the confidence interval is $0.071$ (see Tab. 3. To reduce this length to $0.06$, the sample size should be over $7000$ observations. In practice, collecting such a large number of observations may be costly.
\begin{figure}[H]
	\begin{center}
		\includegraphics[scale=0.5]{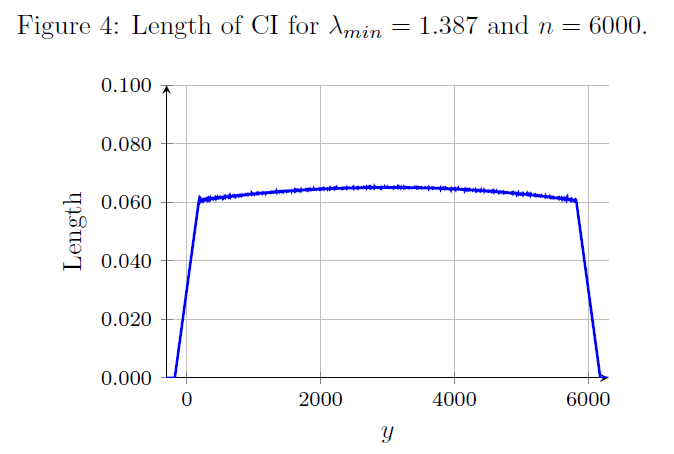}
	\end{center}
\end{figure}
\begin{table}[H]
	\begin{center}
		\includegraphics[scale=0.5]{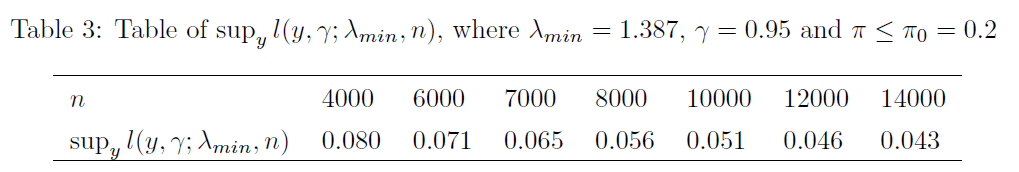}
	\end{center}
\end{table} 

\bigskip
\noindent{\bf 4. Asymptotic confidence interval}
%\label{sec: asymtotic}\input{Sec/asymptotic}

	Let us denote the sample means for the first and second treatment groups by 
${\bar Y}^{(1)}=\frac{1}{n_1}\sum_{i=1}^{n_1}Y_{1i}$ and ${\bar Y}^{(2)}=\frac{1}{n_2}\sum_{i=1}^{n_2}Y_{2i}$, respectively. The unbiased method of moments ($MM$) estimator of the sensitive proportion $\pi$ is:
$$\hat\pi_{MM}=\frac{1}{2}\left({\bar Y}^{(2)}-{\bar Y}^{(1)}\right), \eqno{(1)}$$
with the variance:
$$V(\hat\pi_{MM})=\frac{1}{4}\left(\frac{1}{n_1}+\frac{1}{n_2}\right)\left(\lambda+\pi(1-\pi)\right)$$

It should be noted that the optimal division of the sample is into two groups of equal size.
It is common to apply Central Limit Theorem in case of large sample sizes. This theorem claims that $\hat\pi_{MM}$ is asymptoticaly distributed as $N\left(\pi,V(\hat\pi_{MM})\right)$. Therefore, the asymptotic CI for $\pi$ is obtained as a solution of the inequality
$$\frac{\left(\hat\pi_{MM}-\pi\right)^2}{V(\hat\pi_{MM})}\leq u_{\frac{1+\gamma}{2}}^2$$
with respect to $\pi$:
$$\frac{2\hat\pi_{MM}n+u^2\pm u\sqrt{u^2+4\lambda(u^2+n)+4n\hat\pi_{MM}(1-\hat\pi_{MM})}}{2(n+u^2)}.$$
Here $u= u_{\frac{1+\gamma}{2}}$ stands for the quantile of the standard normal distribution, and $n_1=n_2=n/2$.

The actual coverage probability of the asymptotic CI is shown in Fig. 5.
\begin{figure}[h]
	\centering
		\includegraphics[scale=0.5]{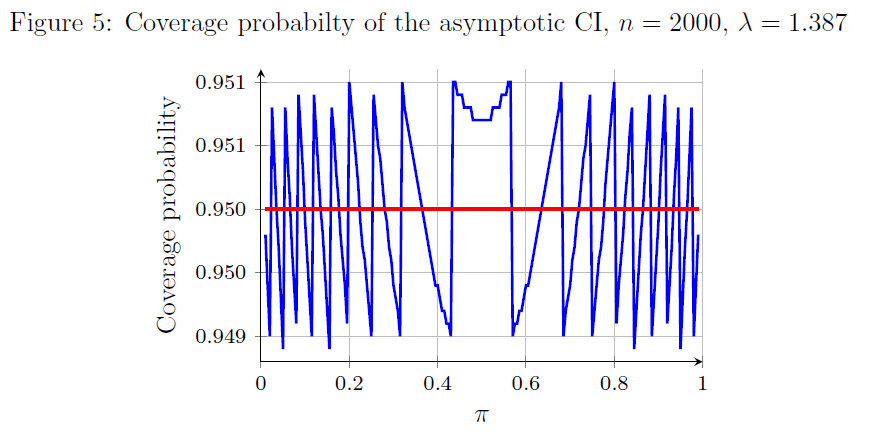}
\end{figure}
The nominal level of confidence is $1-\gamma=0.95$. It is clear that the \textit{actual} confidence level is lower than the nominal one. In contrast with this undesirable behaviour of the asymptotic CI, the exact CI proposed in our paper works quite well even for small samples. 

%	\begin{figure}[H]
	%		\begin{center}
		%			\caption{Length of asymptotic c.i. for $\lambda_{min}=1.387$ and $n/2=1000$.}\label{graph: Asytypicalshape}
		%			%\usebox{\Asytypicalshape}
		%			\usebox{\lengthM}
		%		\end{center}
	%	\end{figure}
Optimal sample sizes for the given criteria in the asymptotic case can be used. 
Let  $\pi_0=0.1$. It implies that $\lambda_{min}=0.693$. Let us compare the case of $n=4900$ and the case of $n=4800$ with respect to the mappings (in this example $l$ denotes the length of the asymptotic CI): 
\begin{enumerate}
	\item $\pi\mapsto E_{\pi}l(Y,\gamma;\lambda_{min},n)$,
	\item $\pi\mapsto P_{\pi}\left\{l(Y,\gamma;\lambda_{min},n)\leq d\right\}$ for $d=0.05$.
\end{enumerate}
The first mapping is shown in Fig. 6 and the second one in Fig. 7. The both mappings are in blue. It is seen in Fig. 6 that according to the first optimality criterion for $d=0.05$ the optimal sample size is equal to at most $4800$ (the blue lines lie below the red line at $0.05$). 

According to the second optimality criterion the size  $n=4800$ is not large enough to hold it. It requires about $50$ observations more. 
%-----------------------

\begin{figure}[H]
	\centering
		\includegraphics[scale=0.5]{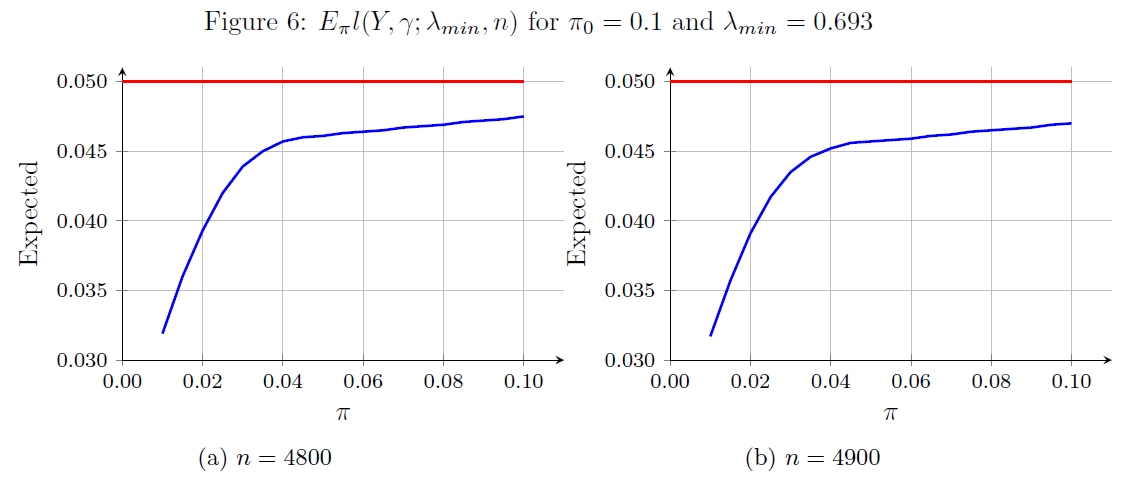}
\end{figure}

\begin{figure}[H]
	\centering
		\includegraphics[scale=0.5]{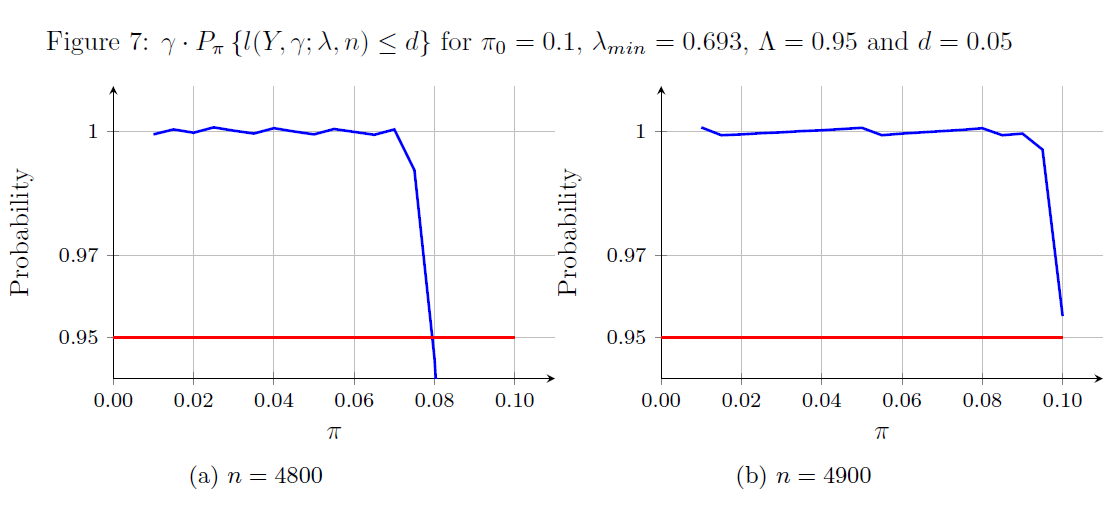}
\end{figure}

\bigskip
\noindent{\bf 5. Summary}

In survey research, we may encounter respondents' reluctance to answer the questions contained in the survey. These are so-called sensitive questions. To ensure the credibility of the research, it is necessary to guarantee the respondents anonymity. Statistical methods that address this issue maintain the respondent's anonymity even when matching the surveyed person to a specific survey. The first such method was proposed by Warner (Warner, 1965). Since then, this method has been modified in several ways, as evidenced by the bibliography included in the article. One of these methods is the ICT.

The original item count technique was introduced by Miller (1984), and since then, various generalizations of the ICT model have been developed to address different scenarios. However, many of these generalizations share a common drawback. Specifically, in an item count design consisting of $K$ non-sensitive questions and one sensitive question, respondents in the treatment group who answer $K+1$ (i.e., respond {\it YES} to all questions) will automatically reveal their sensitive characteristic. This phenomenon is known as the ceiling effect.
To address this drawback, Tian (2014) proposed Poisson and negative binomial count technique models, where the observed variable $Y$ is expressed as 
$$Y=
\begin{cases}
	X ,&\hbox{in the $1^{st}$ sample};\\
	X + Z,&\hbox{in the $2^{nd}$ sample}.\\
\end{cases}
$$
Here, $X$, the count variable, follows either a Poisson or negative binomial distribution.
However, the proposed models still have certain drawbacks. First, reporting zero (as the sum of responses to both neutral and sensitive questions) eliminates any suspicion that the respondent belongs to the sensitive group. Second, the statistical efficiency is quite low, meaning that a very large sample size is required to achieve reasonable precision. To address these issues, Kowalczyk et al. (2021) modified the ICT technique. They proposed to express observed variable $Y$ as  
\begin{equation}\label{keyform}
	Y=
	\begin{cases}
		X -Z,&\hbox{in the $1^{st}$ sample};\\
		X + Z,&\hbox{in the $2^{nd}$ sample},\\
	\end{cases}
\end{equation}
what improves statistical efficiency.
Unfortunately, respondents in the first group who report $-1$ still reveal their sensitive characteristic.

In this article, we chose to retain expression (\ref{keyform}) for the observed variable $Y$, as it improves statistical efficiency. To prioritize privacy protection, we adopted the optimality criteria used by Jaworski and Zieli{\'n}ski (2023).  In doing so, we achieved two key objectives. First, we controlled the occurrence of reporting $-1$ in the first group. Second, we established the appropriate level of privacy protection before determining the sample size. This is crucial, as the sample size required to ensure adequate precision depends significantly on the level of privacy.

The presented approach can be applied to both exact and approximate intervals. Unlike other researchers, we believe that when exact intervals are feasible, approximate intervals should not be used, as they fail to maintain the specified confidence level. Nevertheless, we have also provided examples of applying the optimal sample selection criteria to an approximate confidence interval based on the method of moments. Readers interested in other constructions of approximate confidence intervals are referred to Kowalczyk et al. (2021).

In the paper, we included numerical examples showing that the optimal sample sizes can be relatively large unless additional information about the estimated parameter is used. Therefore, in practical applications, it is advisable to utilize the information that the estimated proportion is small and does not exceed a known value.

\vfill\eject
	
\noindent{\bf References}

	\begin{figure}[H]
		\begin{center}
			\includegraphics[scale=1]{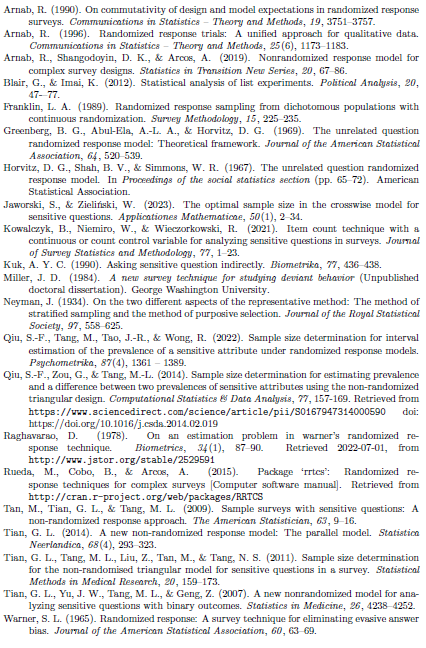}
		\end{center}
	\end{figure}
	
	\vfill\eject
	
	\begin{figure}[H]
		\begin{center}
			\includegraphics[scale=1]{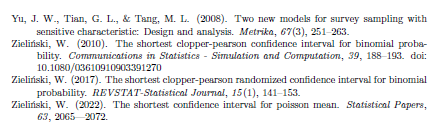}
		\end{center}
	\end{figure}
	
\end{document}